\documentclass[conference]{IEEEtran}
\usepackage{geometry}
\usepackage{graphicx}
\usepackage{subfigure}
\usepackage{pifont}
\usepackage{mathtools}
\usepackage{times}
\usepackage{booktabs} 
\usepackage{multirow}
\usepackage{enumitem}
\usepackage{pifont}
\usepackage{amsmath,amssymb,amsfonts}
\usepackage{algorithmic}
\usepackage{xcolor}
\usepackage{gensymb}
\usepackage{tikz}
\usepackage{xcolor}
\usepackage[bottom]{footmisc}
\newcommand*\circled[1]{\tikz[baseline=(char.base)]{
            \node[shape=circle,fill,inner sep=0.1pt] (char) {\textcolor{white}{#1}};}}

\IEEEoverridecommandlockouts

\begin{document}

\title{\Large\bf MindReading: An Ultra-Low-Power Photonic Accelerator for EEG-based Human Intention Recognition}

\author{
\normalsize
\begin{tabular}[t]{cccccc}
Qian Lou$\S^*$\thanks{$^*$Qian Lou and Wenyang Liu contributed equally. This work was supported in part by NSF CCF-1908992 and CCF-1909509. Wenyang Liu and Weichen Liu were supported by NAP M4082282 and SUG M4082087.}~~~~~~ & Wenyang Liu$\ddag$ & & Weichen Liu$\ddag$~~~~~~ & Feng Guo$\S$ & Lei Jiang$\S$\\
\multicolumn{3}{c}{$\S$Indiana University Bloomington, USA} & \multicolumn{3}{c}{$\ddag$Nanyang Technological University, Singapore}\\
\multicolumn{3}{c}{\{louqian, fengguo, jiang60\}@iu.edu}    & \multicolumn{3}{c}{\{wenyang.liu, liu\}@ntu.edu.sg}\\
\end{tabular}
}

\maketitle

\makeatletter
\def\ps@IEEEtitlepagestyle{%
  \def\@oddfoot{\mycopyrightnotice}%
  \def\@evenfoot{}%
}
\makeatother
\def\mycopyrightnotice{%
  \begin{minipage}{\textwidth}
    \footnotesize
    978-1-7281-4123-7/20/\$31.00~\copyright~2020 IEEE \hfill\\~\\
  \end{minipage}
  \gdef\mycopyrightnotice{}% just in case
}

{\small\bf Abstract--- A scalp-recording electroencephalography (EEG)-based brain-computer interface (BCI) system can greatly improve the quality of life for people who suffer from motor disabilities. Deep neural networks consisting of multiple convolutional, LSTM and fully-connected layers are created to decode EEG signals to maximize the human intention recognition accuracy. However, prior FPGA, ASIC, ReRAM and photonic accelerators cannot maintain sufficient battery lifetime when processing real-time intention recognition. In this paper, we propose an ultra-low-power photonic accelerator, MindReading, for human intention recognition by only low bit-width addition and shift operations. Compared to prior neural network accelerators, to maintain the real-time processing throughput, MindReading reduces the power consumption by 62.7\% and improves the throughput per Watt by 168\%.
}

\section{Introduction}
\label{s:intro}

Brain-computer interface (BCI)~\cite{Machado:RIN2010} enables the direct communications and control using brain intentions alone, and thus offers a practical way to help people suffering from motor disabilities. Particularly, scalp-recording electroencephalography (EEG)~\cite{OpenBCI:O2018,Lazarou:FIHN2018} is one of the most promising solutions to implementing BCIs, due to its low-cost and portable acquisition system. When a person is intent on moving different parts of his body, the EEG signals from his scalp fluctuates in different modes. In this way, human intentions can be recognized by decoding EEG signals. EEG-based BCI has been widely adopted in controlling wheelchairs, prosthetics and exoskeletons~\cite{Simic:FHN2018}.

However, recognizing human intentions by decoding EEG signals is challenging. EEG-based BCI systems suffer from inevitable noises~\cite{Lazarou:FIHN2018}, due to human physiological activities, e.g., eye blinks and heart beats. Moreover, the correlations~\cite{Lazarou:FIHN2018} between EEG signals and their corresponding brain intentions are not straightforward. To denoise EEG signals and detect human intentions, prior works~\cite{zhang:AAAI2018,Chen:ISLPED2018} create neural networks consisting of multiple LSTM and convolutional layers that obtain high recognition accuracy (e.g., 98.3\%~\cite{zhang:AAAI2018}). Because of the $128Hz$ raw EEG signal sampling rate~\cite{zhang:AAAI2018}, to recognize intentions in real time, a BCI system processes the inference of a typical EEG neural network~\cite{zhang:AAAI2018} under the throughput of 128 times per second. For 64-channel EEG signals, the BCI system has to support a $\sim$\textbf{100M}-FLOPS throughput, which is difficult to be delivered by mobile CPUs and GPUs~\cite{Du:ISCA2015} under the tight power constraint and the temperature budget of a $2\degree C$ increase~\cite{Lazzi:EMBM2005} for most bio-embedding applications. The essential computing effect of the EEG-based intention recognition makes mobile CPUs and GPUs~\cite{Du:ISCA2015} hardly meet the real-time processing requirement under the power and temperature constraints.

Although FPGA~\cite{Chen:ISLPED2018}, ASIC~\cite{Du:ISCA2015}, ReRAM~\cite{Shafiee:ISCA2016}, and even photonic~\cite{Liu:DATE2019} neural network accelerators are proposed to process neural network inferences in an energy-efficient way, it is still difficult for the BCI system to adopt these solutions, because of its tight power budget and real-time requirement. The CMOS-based FPGA~\cite{Chen:ISLPED2018} and ASIC~\cite{Du:ISCA2015} designs cannot maintain reasonable battery lifetime when processing neural network inferences. For instance, the battery of Google Glass using an ASIC accelerator stands for only 45 minutes~\cite{Muensterer:IJS2014} when tracking consecutive object actions. The power-hungry CMOS analog-to-digital converters dominate $>80\%$ of the total power consumption of the ReRAM-based accelerator~\cite{Shafiee:ISCA2016} and hence becomes the obstacle to this accelerator's fast adoption in the wearable BCI systems. Inspired by the low power photonic network-on-chip~\cite{Hendry:SC2010}, a recent work~\cite{Liu:DATE2019} creates a photonic accelerator to significantly improve the inference throughput per Watt of convolutional neutral networks by compact optical micro-disks. But the eDRAM and optical adders in the photonic accelerator consume \textbf{79.1}$\%$ of its total power and prevents it from achieving higher power efficiency.

To process the real-time EEG-based human intention recognition more efficiently under tight power and temperature constraints, in this paper, we propose an ultra-low-power photonic accelerator, \textit{MindReading}, for the wearable BCI system. Our contributions can be summarized as follows.

\begin{figure*}[t!]
%\vspace{-0.1in} 
\centering
\includegraphics[width=0.9\textwidth]{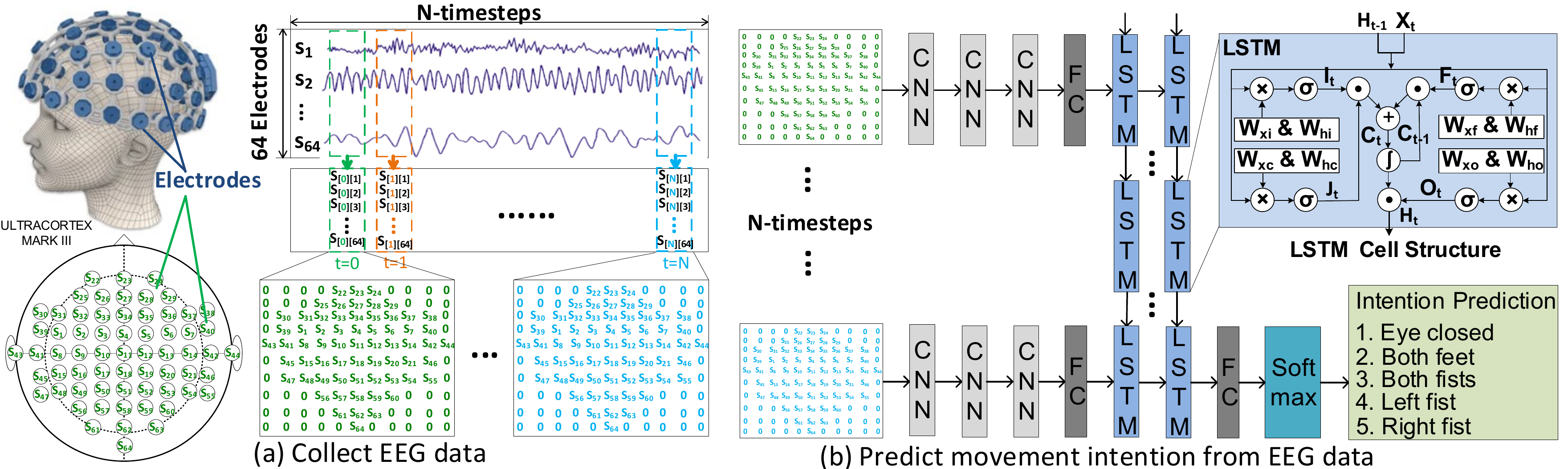}
\vspace{-0.1in}
\caption{The EEG-based Human Intention Recognition.} 
\label{f:eeg_background}
\vspace{-0.2in}
\end{figure*}

\begin{itemize}[noitemsep,topsep=0pt,parsep=0pt,partopsep=0pt]
\item We present universal logarithmic quantization to quantize not only weights but also activations of convolutional, LSTM and fully-connected layers into the data representation of power-of-2 with trivial accuracy degradation. In this way, expensive floating point matrix-vector multiplications can be replaced by low bit-width addition and shift operations.

\item We build a novel photonic human intention accelerator, MindReading, to process the neural network composed of power-of-2 quantized weights and activations by on-chip photonic low-bit adders and shifters. Particularly, we create a photonic activation unit to directly quantize the outputs of various activations, i.e., $Tanh$, $ReLU$ and $Sigmoid$, to power-of-2 representations.

\item We evaluated and compared MindReading against the state-of-the-art CPU, GPU, FPGA, ASIC, ReRAM, photonic neural network accelerators. Our experimental results show that to maintain the real-time processing throughput, MindReading reduces the power consumption by 63\% and improves the throughput per Watt by 170\% over a recent photonic accelerator.
\end{itemize}

\section{Background}

\subsection{Electroencephalography Signal Recognition}
The recognition flow of EEG signals is shown in Figure~\ref{f:eeg_background}. The EEG-based BCI system uses a wearable headset with 64 electrodes to capture EEG signals~\cite{zhang:AAAI2018}. The raw data from 64 electrodes at time-step $t$ is a 1D data vector with the size of 64. For instance, when $t$ is 0, the 1D raw data is $[S_{[0][1]}, S_{[0][2]}, \ldots, S_{[0][64]}]$. To model the position information of electrodes, the 1D raw data vector is converted to a 2D $10\times11$ data matrix according to the 64-electrode placement map shown in Figure~\ref{f:eeg_background}. And then, human intentions can be recognized by decoding EEG signals with high accuracy ($98.3\%$) using EEG-NET~\cite{zhang:AAAI2018} composed of convolutional, fully-connected, LSTM and $soft\,max$ layers. To recognize human intentions in real-time, EEG-NET has to process 128 2D data matrices per second, since the EEG sampling rate of the BCI system is 128Hz~\cite{zhang:AAAI2018}. To reliably adopt a battery-powered real-time BCI system~\cite{Machado:RIN2010,OpenBCI:O2018,Lazarou:FIHN2018} in real-world applications, a low-power human intention recognition hardware accelerator becomes a must.

\begin{figure}[hbtp!]
\vspace{-0.1in} 
\centering
\includegraphics[width=3.5in]{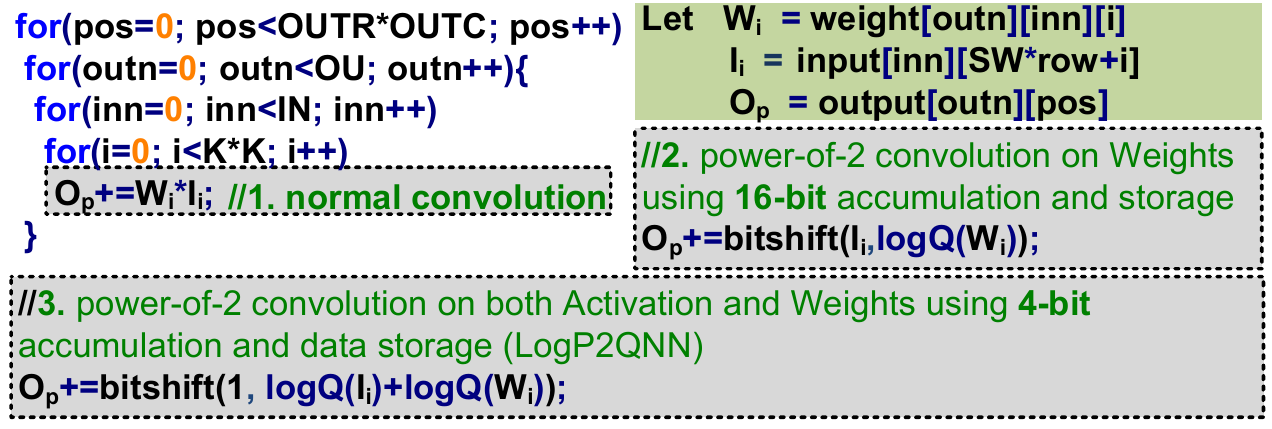}
\vspace{-0.25in}  
\caption{A P2QNN or LogP2QNN quantized convectional layer.} 
\label{f:cnn_pseudo_code}
\vspace{-0.25in}  
\end{figure}

\subsection{Convolutional Layer}
As Figure~\ref{f:cnn_pseudo_code} shows, a convolutional layer takes \textit{IN} $\times$ \textit{INC} $\times$ \textit{INR} as input where \textit{IN}, \textit{INC} and \textit{INR} indicate the input channel number, input width and height, respectively. A \textit{IN} $\times$ \textit{K} $\times$ \textit{K} weight filter convolves with the input by moving \textit{SW} strides until generating \textit{OU} $\times$ \textit{OUTC} $\times$ \textit{OUTR} output elements where \textit{K} is the filter size; \textit{OU}, \textit{OUTC}, and \textit{OUTR} denote the output channel number, width and height, respectively.

\subsection{Long Short-Term Memory Layer}
Figure~\ref{f:eeg_background}(b) shows the basic structure of a Long Short-Term Memory (LSTM) cell, where $H_{t}$ is the output of the time-step $t$, $X_t$ means the input of the time-step $t$; and $C_t$ indicates the cell memory storage. The cell's state and its output are updated by four gates, i.e., $I_t$, $F_t$, $J_t$ and $O_t$. The activation functions ($\sigma$), ($\int$) are \emph{$Sigmoid$} and \emph{$Tanh$}, respectively. And $\bigotimes$, $\bigodot$ and $\bigoplus$ indicate dot-product, element-wise multiplication and element-wise addition, respectively.

\subsection{Logarithmic Quantization}
To reduce the computing overhead, Power-of-2 Quantized Neural Network (P2QNN)~\cite{Liu:DATE2019,E.H:ICASSP2017} is proposed to quantize weights of convolutional layers to their power-of-2 representations. In this way, expensive multiplications can be replaced by cheap binary shift and linear accumulation operations. As Figure~\ref{f:cnn_pseudo_code} shows, P2QNN linearly accumulates 16-bit fixed point inputs to compute a convolutional layer. To further reduce the accumulation overhead, the logarithmically accumulated P2QNN (LogP2QNN)~\cite{E.H:ICASSP2017} is presented by quantizing inputs, weights and even the activations of convolutional layers to their power-of-2 data representations. In Figure~\ref{f:cnn_pseudo_code}, the logarithmic accumulations can be done by lower bit-width (e.g., 4-bit) adders, indicating lower power consumption. Compared to the full-precision model, LogP2QNN decreases the inference accuracy by $\sim1\%$~\cite{E.H:ICASSP2017}. However, applying LogP2QNN on LSTM layers is not trivial, since compared to convolutional layers relying only on $ReLU$, they have more types of activation function including \emph{$Sigmoid$} and $Tanh$. In this paper, we propose an universal logarithmic quantization to quantize activations of LSTM layers with little accuracy degradation.

%To reduce the computing overhead, Power-of-2 Quantized Neural Network (P2QNN)~\cite{Liu:DATE2019,E.H:ICASSP2017} is proposed to quantize weights of convolutional layers to their power-of-2 representations. In this way, expensive multiplications can be replaced by cheap binary shift and linear accumulation operations. As Figure~\ref{f:cnn_pseudo_code} shows, P2QNN linearly accumulates 16-bit fixed point inputs to compute a convolutional layer. To further reduce the accumulation overhead, the logarithmically accumulated P2QNN (LogP2QNN)~\cite{E.H:ICASSP2017} is presented by not only quantizing inputs, weights and even the activations of convolutional layers to their power-of-2 data representations, but also accumulating the multiplication of weights and inputs/activations in log-domain. In Figure~\ref{f:cnn_pseudo_code}, the logarithmic accumulations can be done by lower bit-width (e.g., 4-bit) adders, indicating lower power consumption. Compared to the full-precision model, LogP2QNN decreases the inference accuracy by $\sim1\%$~\cite{E.H:ICASSP2017}. However, applying LogP2QNN on LSTM layers is not trivial, since compared to convolutional layers relying only on $ReLU$, they have more types of activation function including \emph{$Sigmoid$} and $Tanh$. In this paper, we propose an universal logarithmic quantization to quantize activations of LSTM layers with little accuracy degradation.

\begin{figure}[ht!]
\vspace{-0.1in} 
\centering
\includegraphics[width=3.4in]{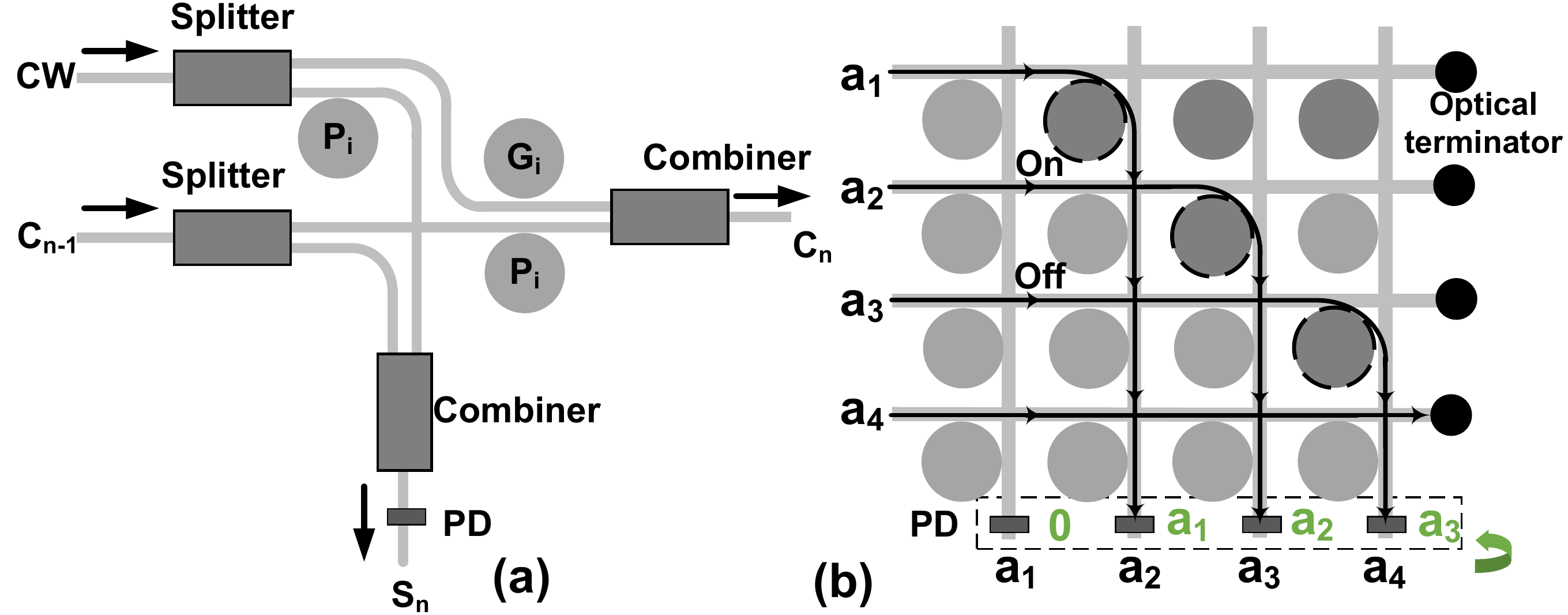}
\vspace{-0.1in}
\caption{Micro-disk-based (a) 1-bit EO full adder (b) 4-bit crossbar shifter.} 
\label{f:adder_shifter}
\vspace{-0.2in} 
\end{figure}

\subsection{Photonic P2QNN Accelerator}

A recent work~\cite{Liu:DATE2019} proposes a photonic accelerator, \textit{Holy-Light-A}, to process P2QNN quantized inferences by micro-disk-based adders and shifters. It achieves the state-of-the-art inference throughput per Watt, since micro-disks have ultra-low power consumption, and high switching frequency.

HolyLight-A adopts a 16-bit ripple-carry adder consisting of 16 1-bit full adders, each of which can be viewed in Figure~\ref{f:adder_shifter}(a). To perform a $N$-bit addition of $A+B$, the carry ($C_i$) and sum ($S_i$) bit calculation are summarized as $C_i = (A_i \oplus B_i) \cdot C_{i-1} + A_i \cdot B_i = P_i \cdot C_{i-1} + G_i$ and $S_i = C_{i-1} \oplus (A_i \oplus B_i) = C_{i-1} \oplus P_i$, respectively, where $i$ means the $i_{th}$ bit. Because the critical path of an $N$-bit carry-ripple adder is determined by the sequential carry bit calculation, so only the carry bit calculation is implemented by photonic micro-disks, while the other parts, i.e., $P_i \& G_i$, are caculated by CMOS transistors~\cite{ying2018silicon} ($\sim 10ps$). Two carrier waves (CWs) are injected to a full adder. Only a CW carries the signal $C_{i-1}$. Both CWs are divided into half by splitters. The electrically computed signals $G_i$ and $P_i$ are applied on micro-disks to modulate the passing lights. By tuning the phase and intensity~\cite{ying2018silicon}, one optical combiner is served as an XOR gate to produce the sum bit, while the other is used as an OR gate to generate the carry bit. The 16-bit adder performance is mainly decided by the modulation speed of micro-disks on the critical path. When micro-disks run at $5GHz$, a 16-bit adder can be reliably operated at $4.3GHz$.

For shift operations, HolyLight-A uses a crossbar composed of $16 \times 16$ micro-disk-based crossing switching elements (CSEs). Figure~\ref{f:adder_shifter}(b) shows a 4-bit crossbar doing a 1-bit logical right shift operation. By configuring the ON or OFF state of the micro-disk, the passing light can turn its direction by 90 degrees. A 4-bit crossbar can implement any $i$-bit right/left binary shift operation by configuring the micro-disk states in the crossbar. If no light is detected by a photodetecter (PD), the output (e.g., $a_1$) is 0. The frequency of a 16-bit shifter is decided by the micro-disk switching speed (4.3GHz).

\begin{figure}[t!]
\centering
\subfigure[HolyLight-A's power breakdown.]{
\includegraphics[width=1.62in]{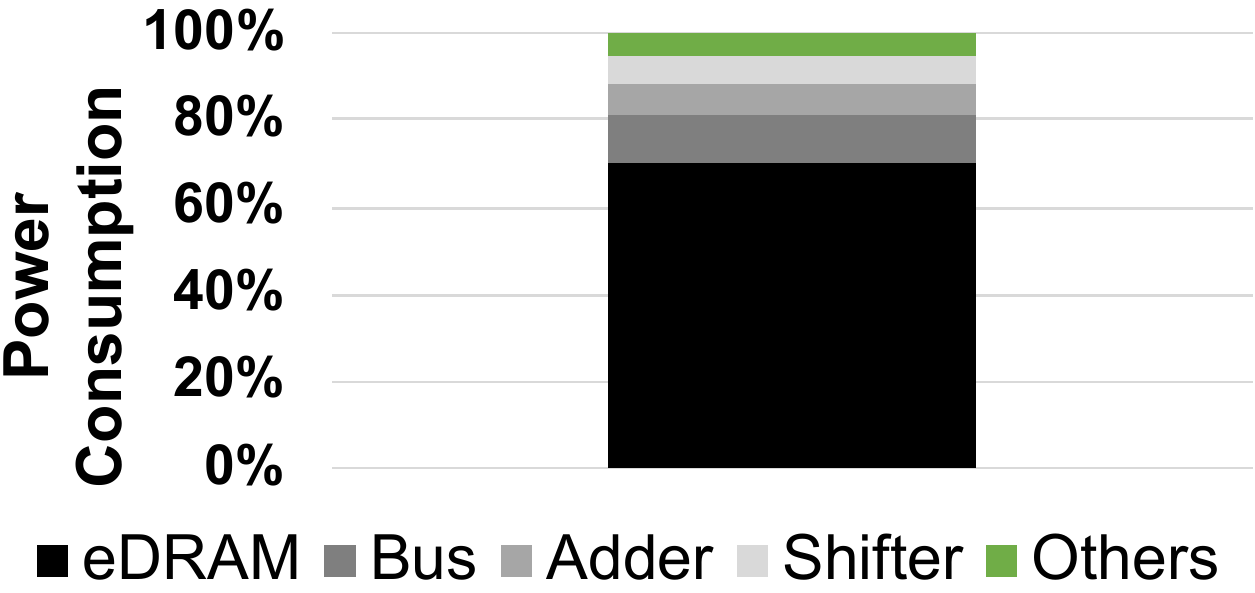}
\label{f:holy_power_analysis}
}
\subfigure[The performance comparison.]{
\includegraphics[width=1.55in]{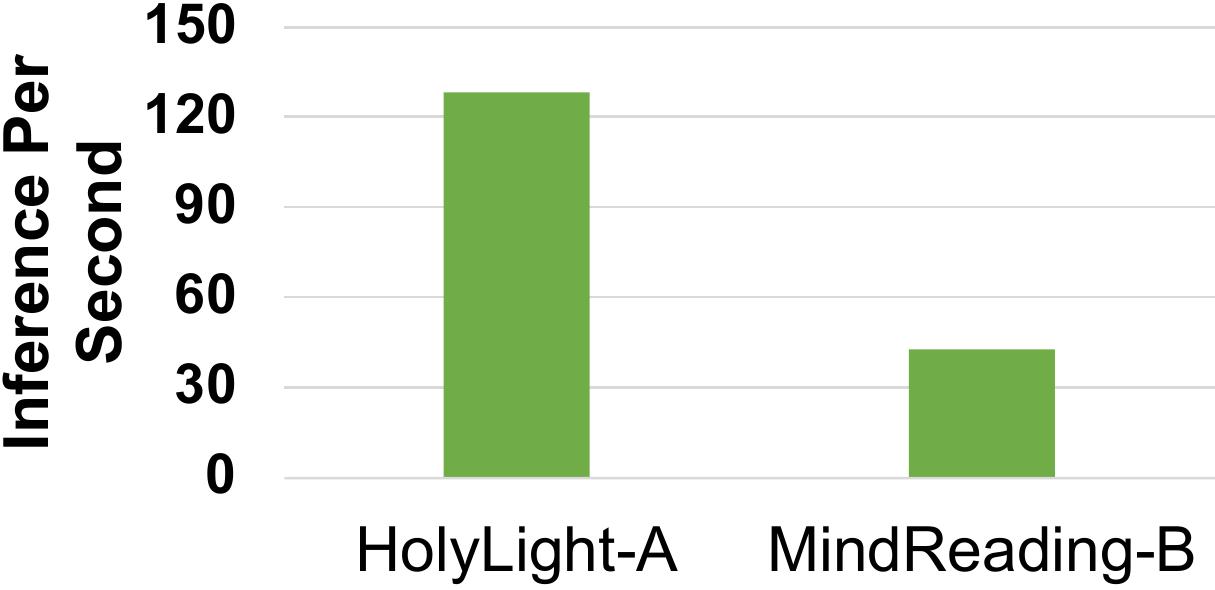}
\label{f:FPS_same_resource}
}
\vspace{-0.1in}
\caption{The power bottleneck of HolyLight-A when accelerating EEG-NET to recognize human intentions in real-time.}
\label{f:bott2}
\vspace{-0.2in}
\end{figure}

\section{Motivation}
%\vspace{-0.1in}

To achieve the real-time processing throughput, a human intention recognition accelerator needs to perform 128 EEG-NET inferences per second (IPS), since the EEG sampling rate of the BCI system is 128Hz~\cite{zhang:AAAI2018}. We customize the original HolyLight-A to a low-power real-time configuration shown in Table~\ref{t:t_power_area} by reducing the unnecessary computing components and lowering the operating frequency. More details can be seen in Section~\ref{s:mindreading}-\ref{s:low_real_cust}. As Figure~\ref{f:FPS_same_resource} shows, the customized HolyLight-A can achieve exactly 128 IPS when processing P2QNN quantized EEG-NET. However, the power consumption of the customized HolyLight-A is still significant for a battery-powered real-time BCI system, due to its power hungry eDRAM buffer, bus, and 16-bit photonic adder. As Figure~\ref{f:holy_power_analysis} shows, in the customized HolyLight-A, the eDRAM, bus and adder consume 71.7\%, 12.1\% and 7\% of its power consumption, respectively. The adder is used for 16-bit accumulations, while the bus and eDRAM are used to transfer and store 16-bit accumulated intermediate results.

To further reduce the power consumption but maintain the same real-time processing throughput, from the \textit{algorithm} perspective, we propose universal logarithmic quantization to quantize both activations and weights for convolutional, LSTM, and fully connected layers in EEG-NET, so that we can replace the 16-bit accumulations by cheaper 4-bit accumulations with little accuracy degradation. From the \textit{hardware} perspective, we present a photonic accelerator to process the neural network composed of power-of-2 quantized weights and activations by on-chip photonic low-bit adders and shifters.

\begin{figure}[htbp!]
\centering
\includegraphics[width=3.3in]{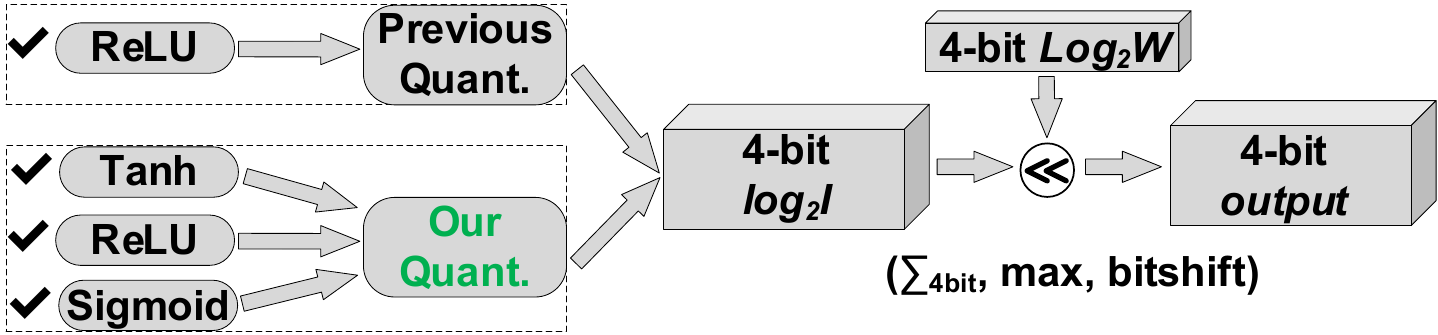}
\vspace{-0.15in}
\caption{Universal Logarithmic Quantization for EEG-NET.} 
\label{f:eeg_quantize}
\vspace{-0.25in}  
\end{figure}

\section{MindReading}
\label{s:mindreading}

\subsection{Universal Logarithmic Quantization}
Since the quantization of LogP2QNN~\cite{E.H:ICASSP2017} is intended for CNNs that only have $ReLU$ activations, we cannot simply apply it on EEG-NET that includes other types of activations, e.g., $Tanh$ and $Sigmoid$. As Figure~\ref{f:eeg_quantize} shows, we propose an universal logarithmic quantization (ULQ) method to quantize \emph{$Sigmoid$}, $Tanh$ and $ReLU$ activations to the power-of-2 representations. The ULQ adopts the \textit{same} method as LogP2QNN~\cite{E.H:ICASSP2017} to quantize weights.

\vspace{-0.2in}
\begin{small}
\begin{gather}
TanhLogQuant(I, N)=sign(I)\times 2^{\overline{I}}\label{e:tanh0}\\
\overline{I}=\begin{cases}
0 & \text{if $I$ = 0},\\
Clip(Round(Log_2|I|,\alpha-N, \alpha)& \text{if $I$ $\neq$ 0}.
\end{cases} \label{e:tanh1}
\end{gather}
\end{small}
\vspace{-0.3in}

\begin{small}
\begin{gather}
Clip(a,min,max)=\begin{cases}
a & \text{if $a$ $\in$ [$min$, $max$]},\\
min& \text{if $a$ $<$ $min$},\\
max& \text{if $a$ $>$ $max$}.\\
\end{cases}\label{e:clip}
\end{gather}
\end{small}
\vspace{-0.1in}

\begin{figure*}[t!]
\centering
\includegraphics[width=6.8in]{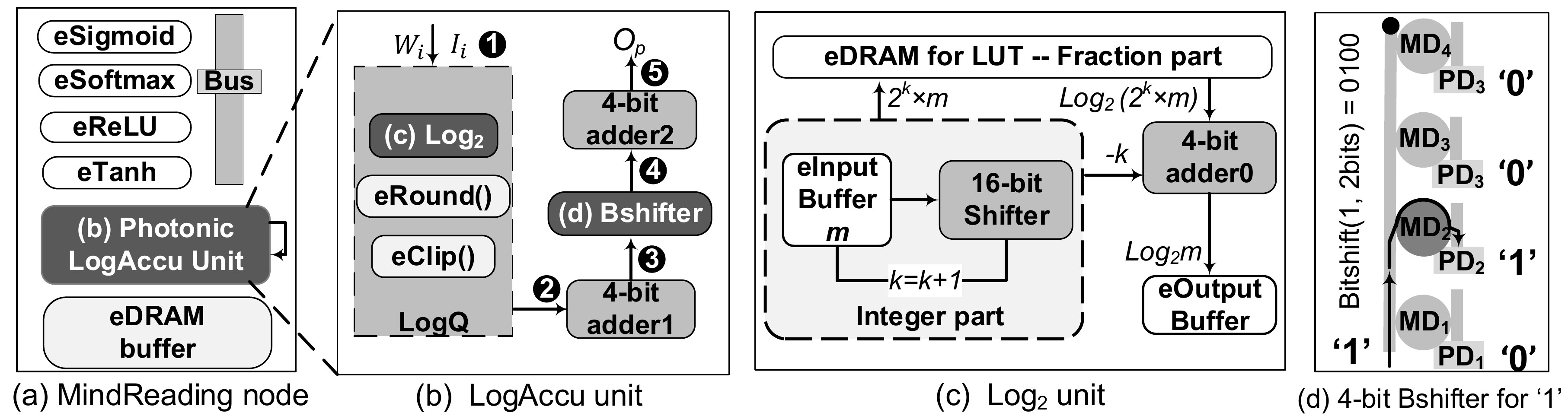}
\vspace{-0.1in} 
\caption{The architecture and pipeline of MindReading.} 
\label{f:architecture}
\vspace{-0.2in}  
\end{figure*}

As Equation~\ref{e:tanh0} and \ref{e:tanh1} show, we present the ULQ function $TanhLogQuant(I,N,is\_Tanh)$ to quantize a $Tanh$ activation to an $N$-bit power-of-2 representation. Particularly, in Equation~\ref{e:tanh1}, the function of $Clip(a, min, max)$ (explained by Equation~\ref{e:clip}) clips the input $a$ to the range [$min$, $max$]. The function of $Rounds(a)$ bounds the input $a$ to the closest integer. The range of $Tanh$ values is $(-1,1)$, so the $min$ and $max$ values in the $clip()$ function are $-N$ and $0$, respectively. The constant $\alpha$ controls the offset range of ULQ and its default value is 0. Through changing $\alpha$, we can fine-tune the range of the quantized $Tanh$ activation value to obtain higher inference accuracy during training.

Similarly, to quantize a $Sigmoid$ activation, we can use the ULQ described in Equation~\ref{e:Sigmoid0} and \ref{e:Sigmoid1}. The $Sigmoid$ activations fall in the range of $(0,1)$. The $min$ and $max$ values in the $clip()$ function for $Sigmoid$ activations are $\beta-N$ and $\beta$, respectively. $\beta$ decides the range of quantized $Sigmoid$ activations. We set the default $\beta$ value as $1$.

\vspace{-0.2in}
\begin{small}
\begin{gather}
SigmoidLogQuant(I, N)= 2^{\overline{I}}\label{e:Sigmoid0}\\
\overline{I}=Clip(Round(Log_{2}|I|),\beta-N,\beta)\label{e:Sigmoid1}
\end{gather}
\end{small}
\vspace{-0.2in}
 
To quantize a non-negative $ReLU$ activation, we can adopt the ULQ in Equation~\ref{e:relu0}. Since the range of $ReLU(x)$ is in $[0,x)$ and the distribution of $ReLU$ is different from those of $Sigmoid$ and $Tanh$, its $\overline{I}$ can be computed by Equation~\ref{e:relu1}. The default $\theta$ value is 0. 

\vspace{-0.2in}
\begin{small}
\begin{gather}
ReLULogQuant(I, N)= 2^{\overline{I}}\label{e:relu0}\\
\overline{I}=Clip(Round(Log_{2}|I|),\theta,\theta+N)\label{e:relu1}
\end{gather}
\end{small}
\vspace{-0.2in}

In short, our proposed ULQ can quantize $Tanh$, $ReLU$ and $Sigmoid$ activations to power-of-2 representations with negligible accuracy loss. Specifically, 4-bit ULQ-quantized EEG-NET has 97.6\% accuracy, degrading the inference accuracy by only 0.7\% over the full-precision EEG-NET.

\subsection{MindReading Photonic Accelerator}
\subsubsection{Architecture}

The overall architecture of MindReading is shown in Figure~\ref{f:architecture}. The chip node relies on an eDRAM buffer to store EEG signals and intermediate results generated by Photonic Processing Unit (LogAccu unit). The LogAccu unit is responsible to calculate binary logarithms and logarithmic accumulations of ULQ-quantized EEG-NET  mainly by using photonic adders and shifters. The chip node adopts electrical nonlinear units for EEG-NET activations including $ReLU$, $Tanh$ and $Sigmoid$.

\subsubsection{MindReading LogAccu Unit}

As Figure~\ref{f:architecture}(b) shows, the MindReading LogAccu unit is in charge of processing the convolutional, LSTM and fully-connected layers of ULQ-quantized EEG-NET. The weights are quantized during training and can be fetched to eDRAMs. The EEG input signals and activations are quantized at run-time by ULQ. During EEG-NET inferences, inputs/activations and quantized weights are read from the input buffer and allocated to the \textit{LogAccu} unit. The inputs/activations are ULQ-quantized by a photonic $Log_2$ unit. And then, two 4-bit photonic adders and a Bshifter in the \textit{LogAccu} unit collaboratively compute the accumulations in logarithmic domain. The intermediate results of the \textit{LogAccu} unit are cached in an output buffer for the next-layer processing.

\textbf{LogAccu unit Components}. We implement each component of the MindReading LogAccu unit as follows:

\begin{itemize}[noitemsep,topsep=0pt,parsep=0pt,partopsep=0pt]

\item \textbf{Photonic $Log_2$ unit}. We build a photonic $Log_2$ unit shown in Figure~\ref{f:architecture}(c) to accelerate binary logarithm computations. $Log_2(m)= Log_2{(2^{-k} \times 2^k \times m)} = -k + Log_2{(2^k \times m)}$, where $m$ is inputs/activations and weights, and mapped into $(1,2]$ by multiplying $2^{k}$ using a photonic shifter, so that $-k$, $Log_2{(2^k \times m)}$ are the integer part and fraction part of $Log_2(m)$. The integer part, $-k$, is determined by checking the result after each 1-bit shift until $m$ is mapped into $(1,2]$. Since outputs of each layer are normalized into the range of (-1,1) by the non-linear activation functions, e.g. $Sigmoid$, $Tanh$, the integer part $-k$ can be determined in one cycle. The fraction part is returned by searching a tiny look-up table ($\sim 8KB$) in eDRAM storing the $log_2$ values between $(1,2]$. Finally, two parts are summed to obtain $Log_2(m)$ using a 4-bit photonic adder.

\item \textbf{eRound and eClip}. We use CMOS $eRound$ and $eClip$ units to facilitate a photonic $Log_2$ unit to construct the ULQ-quantization LogQ unit, where the $Log_2$ computation is the most time-consuming step.

\item \textbf{Photonic 4-bit Adder}: We adopt the same photonic ripple carry adder design from HolyLight-A~\cite{Liu:DATE2019}. 

\item \textbf{Photonic 4-bit Bshifter}. To compute $bitshift(1,B)$, we propose a low-cost photonic 4-bit Bshifter shown in Figure~\ref{f:architecture}(d) by micro-disk-based parallel switching elements (PSEs). As Figure~\ref{f:cnn_pseudo_code}3 shows, LogP2QNN only requires the values of $bitshift(1,B)$ during convolutions. Hence a general photonic 4-bit shifter is not considered for saving the power and energy. In addition, both PSEs and CSEs can change the direction of waves, but PSEs have a more compact size and less insertion loss. Our ULQ also shares the same principle to process convolutional, LSTM and fully-connected layers. By configuring the MDs into ON or OFF states, Bshifter can shift the input 1 by $B$ bits. Figure~\ref{f:architecture}(d) shows an example of $Bitshift(1,2)$, where the second MD, $MD_2$, is set to ON state.
\end{itemize}

\textbf{LogAccu Pipeline}. 
To implement ULQ quantization, shift and accumulation operations, LogAccu unit requires 9 cycles to derive $O_p$ from weight $W_i$ and input/activation $I_i$. As Figure~\ref{f:architecture}(b) describes, \circled{1} $W_i$ and $I_i$ are fetched from eDRAM buffer using one cycle. \circled{2} 5 cycles are required to calculate $LogQ(I_i)$ and $LogQ(W_i)$. These 5 cycles are for integer part computation, fraction part computation, sum between those tow parts in $Log_2$ unit, $eClip()$ and $eRound()$, respectively. \circled{3} In the 7th cycle, the sum $LogQ(I_i)+LogQ(W_i)$ is calculated. \circled{4} $Bshifter$ outputs bitshift$(1, LogQ(I_i)+LogQ(W_i))$ in the 8th cycle, meanwhile, the last time-step of $O_{p}$ is loaded from eDRAM buffer. \circled{5} 4-bit adder2 sums the last time-step $O_{p}$ and bitshift$(1, LogQ(I_i)+LogQ(W_i))$ in the 9th cycle. The accumulation using 9 cycles will be constantly performed until one entire convolutional result, $O_p$ , is generated. After that, the generated $O_p$ will be be activated using activation functions, e.g. $ReLU$ and $Tanh$, for the next-layer processing. The loop of accumulation in log-domain and activation won't stop until the entire EEG-NET inference is finished.

\subsubsection{Low Power Real-time Hardware Customization}
\label{s:low_real_cust}
The design goal of the human intention recognition accelerator is to minimize the power consumption while maintaining a 128 IPS throughput. To use HolyLight-A to process EEG-NET, we scaled its frequency down and adjusted the number of its hardware resources, e.g., photonic adders and shifters. We found that one 16-bit adder and one shifter operating at $4.3GHz$ are enough to make HolyLight-A to achieve the real-time processing throughput of EEG-NET. We call it the customized HolyLight-A. We construct the baseline of MindReading (MindReading-B) by one 4-bit adder and a shifter operating at $4.3GHz$. As Figure~\ref{f:FPS_same_resource} shows, unfortunately, MindReading-B obtains only 43 IPS, indicating it cannot meet the real-time requirement. To enable MindReading to achieve 128 IPS, we add another two 4-bit adders in MindReading-B.

%Since HolyLight-A is built for data center, to meet up with both low power and the real-time processing requirement of EEG system, we just simply scale down the frequency and the number of basic photonic devices, i.e., adders and shifters. More details about scaling can be found in  Section~\ref{s:expmeth}. One 16-bit adder and shifter at 4.3Ghz is enough to implement P2QNN's basic operations under the real-time processing requirement of EEG system. To reduce HolyLight-A's power consumption, we simply use one 4-bit adder and shifter under the same frequency (4.3Ghz) with HolyLight-A, to construct MindReading's baseline design (MindReading-B). However, this MindReading baseline just predicts 43 2D EEG signals in one Second shown in Figure~\ref{f:FPS_same_resource} which can't meet the minimum real-time processing requirement, 128 FPS, although MindReading's baseline reduce $\sim$4-times less power than HolyLight-A. The reason why MindReading's baseline suffers from low performance is that a 4-bit adder and 16-bit adder under 4.3Ghz actually share the same latency but LogP2QNN has 3-times additions than P2QNN. To solve this problem, we add two 4.3Ghz 4-bit adders into MindReading baseline to construct our final-version MindReading.

\begin{table}[htbp!]
\vspace{-0.1in}
\centering
\scriptsize
\caption{The power and area comparison between MindReading and HolyLight-A.}
\vspace{-0.1in}
\setlength{\tabcolsep}{0.88pt}
\begin{tabular}{|c||c|c|c|c|}\hline
Name                          & Component            & Spec                    & Power ($mW$)  & Area ($mm^2$)  \\\hline\hline
                              & 16-bit adder         & $\times 1$, 4.3GHz      & 4.24	       & 0.00788    \\\cline{2-5}	
                              & 16-bit shifter       & $\times 1$, 4.3GHz      & 3.51          & 0.02796   \\\cline{2-5}	                              
                              & eDRAM                & 256KB                   & 41.4	       & 0.16600   \\\cline{2-5}
            4.3GHz 	          & bus                  & 384-wire                & 7             & 0.00900      \\\cline{2-5}
  			HolyLight-A 	  & $eActivation$              & $\times 4$        & 1.04    	   & 0.00120     \\\cline{2-5}
                              & $eClip$              & $\times 1$              & 0.26          & 0.00030    \\\cline{2-5}
                              & $eRound$             & $\times 1$              & 0.26          & 0.00030    \\\cline{1-5}
Total                         &                      &                         & 57.71         & 0.21264    \\\hline\hline

                              & Bshifter             & $\times 1$, 4.3GHz      & 0.87	       & 0.00024    \\\cline{2-5}	
                              & 16-bit shifter       & $\times 1$, 4.3GHz      & 3.51          & 0.02796   \\\cline{2-5}
                              & 4-bit adder          & $\times 3$, 4.3GHz      & 2.93          & 0.00591   \\\cline{2-5}
                              & eDRAM                & 64KB                    & 10.4	       & 0.04150   \\\cline{2-5}
    4.3GHz                    & bus                  & 128-wire                & 2.33          & 0.00300     \\\cline{2-5}
  	MindReading			      & $eActivation$              & $\times 4$        & 1.04    	   & 0.00120     \\\cline{2-5}
                              & $eClip$              & $\times 1$              & 0.26          & 0.00030    \\\cline{2-5}
                              & $eRound$             & $\times 1$              & 0.26          & 0.00030    \\\cline{1-5}
Total                         &                      &                         & 21.55         & 0.08041    \\\hline

\end{tabular}
\label{t:t_power_area}
\vspace{-0.1in}
\end{table}

\subsubsection{Design overhead}
The comparison of power and area between of customized HolyLight-A and MindReading are summarized as Table~\ref{t:t_power_area}. HolyLight-A and MindReading share the same electrical activation devices, but they have different sizes of eDRAM buffer. This is because both weights and activations of MindReading are only 4-bit. eActivation represents $eReLU$, $eSigmoid$, $eSoftmax$, or $eTanh$. All electrical logic units are modeled and estimated through Cadence Virtuoso with $32nm$ PTM technology. CACTI is used to model eDRAM, input and output buffers. Similar to HolyLight-A, MindReading uses one photonic I/O~\cite{Liu:DATE2019} to communicate with CPUs. We used Lumerical FDTD~\cite{FDTD} to simulate photonic micro-disk-based computing components. To build MindReading, we modeled and adopted optical splitters \& combiners, photodetectors and micro-disks from~\cite{Liu:DATE2019}. To estimate the MindReading area, we used a systematic analysis tool, CLAP~\cite{duong:VLSI2016}, that provides detailed structures of various optical devices.

\section{Experiment Methodology}
\label{s:expmeth}

\textbf{Workload}. MindReading recognizes human intentions by accelerating EEG-NET~\cite{zhang:AAAI2018} with ultra-low power. We trained EEG-NET with PhysioNet EEG Dataset~\cite{PhysioNet} using PyTorch-v0.4. EEG-NET consists of 3 convolutional, 2 fully-connected, 2 LSTM with 30 time-steps and 1 softmax layers. More EEG-NET details can be viewed in Table~\ref{t:t_eeg_topo}. Compared to the full-precision EEG-NET with accuracy 98.3\%, the ULQ-quantized EEG-NET degrades only 0.7\% inference accuracy.

\begin{table}[htbp!]
\centering
\scriptsize
\caption{The EEG-NET architecture({\scriptsize Conv: convolutional; FC: fully-connected;})}
\vspace{-0.1in}
\setlength{\tabcolsep}{3pt}
\begin{tabular}{|c||c|c|c|c|c|}
\hline
Layer          & Output Size      & Ksize & stride      & Output Channels\\
\hline\hline
Conv1          & 10$\times$11     & 3$\times$3 & 1      & 32\\
\hline
Conv2          & 10$\times$11     & 3$\times$3 & 1      & 64\\
\hline
Conv3          & 10$\times$11     & 3$\times$3 & 1      & 128\\
\hline
FC1            & 1$\times$1       & / & /               & 1024\\
\hline
LSTM1          & 1$\times$1       & / & /               & 64\\
\hline
LSTM2          &  1$\times$1      & / & /               & 64\\
\hline
FC2            &  1$\times$1      &  /& /               & 1024\\
\hline
Softmax        &  1$\times$1      &/  &  /              & 6\\
\hline
\end{tabular}
\vspace{-0.1in}
\label{t:t_eeg_topo}
%\vspace{-0.05in}
\end{table}

\textbf{Accelerators}. We compared MindReading against 7 counterparts shown in Table~\ref{t:t_scheme_comp}. We selected an ARM Cortex-A15 CPU, an Nvidia Tegra-4 GPU, a Zynq-7030 FPGA~\cite{Chen:ISLPED2018}, a ShiDianNao ASIC~\cite{Chen:MICRO2014}, a ReRAM-based CNN accelerator ISAAC~\cite{Shafiee:ISCA2016}, a ASIC binary CNN accelerator MXBCNN~\cite{Bankman:ISSCC2018}, and a photonic CNN accelerator HolyLight-A~\cite{Liu:DATE2019}. ShiDianNao reduces DRAM accesses for weights to speedup deep neural networks. ISAAC relies on ReRAM-based dot-product engines to accelerate matrix-vector multiplications. MXBCNN using XNOR and Popcount engines to accelerate binarized CNN. HolyLight-A depends on photonic adders and shifts to perform P2QNN inferences. The inference accuracy comparison of all accelerators is also shown in Table~\ref{t:t_scheme_comp}. CPU, GPU, FPGA, ShiDianNao and ISAAC implement 16-bit fixed-point EEG-NET with 98.3\% accuracy. MXBCNN degrades 2.2\% accuracy due to its 4-bit binarized weights and activations. HolyLight-A achieves 97.6\% accuracy using 16-bit P2QNN. Although ULQ further quantizes all activations, MindReading still obtains 97.6\% accuracy by 4-bit ULQ. 

\textbf{Customized accelerator configurations}. Since the EEG sampling rate of the BCI system is 128Hz, we customized a real-time configuration that can achieve 128 IPS for each accelerator. Except HolyLight-A and MindReading, we assume the frequency and the number of hardware resources in the other accelerators can be ideally and linearly scaled, so that all accelerators can achieve exactly 128 IPS, e.g., 37.2$\times$ 6W Nvidia Tegra-4 GPU has a 128-IPS throughput. The linear scaling actually overestimates the throughput per Watt of these accelerators, since in most cases their peripheral circuits, e.g., I/O and buses, are not modular or scalable.

%actually can't be scaled down and should be rounded up. 
%A more detail example about the customized method is shown in  Table~\ref{t:t_power_area} which lists the hardware configuration of HolyLight-A after scaling down.    
%We customized all accelerators to their hardware configuration

\begin{table}[htbp!]
\vspace{-0.1in}
\centering
\scriptsize
\caption{Simulated scheme comparison.}
\vspace{-0.1in}
\setlength{\tabcolsep}{3pt}
\begin{tabular}{|c||c|c|}
\hline
Name                                         & Description              & Accuracy ($\%$)\\
\hline\hline
CPU                                          & ARM Cortex-A15           & $98.3$       \\\hline
GPU                                          & Nvidia Tegra 4         & $98.3$      \\\hline
FPGA~\cite{Chen:ISLPED2018}                  & Zynq-7030                & $98.3$    \\\hline
ShiDianNao~\cite{Du:ISCA2015}                & ASIC              & $98.3$    \\\hline
ISAAC~\cite{Shafiee:ISCA2016}                & ReRAM PIM          & $98.3$    \\\hline
MXBCNN~\cite{Bankman:ISSCC2018}              & Binary CNN             & $96.1$    \\\hline
HolyLight-A~\cite{Liu:DATE2019}              & Photonic P2QNN	            & $97.9$    \\\hline
MindReading                                  & Photonic ULQ	            & $97.6$    \\\hline
\end{tabular}
\vspace{-0.1in}
\label{t:t_scheme_comp}
%\vspace{-0.05in}
\end{table}

\textbf{Accelerator modeling}. 
A heavily modified deep learning accelerator simulator FODLAM~\cite{Sampson:FODLAM2018} is used to study the accelerator performance and power. FODLAM has been correlated and validated by physical accelerator chips such as ShiDianNao. Based on a user-defined accelerator configuration and EEG-NET, it can generate the performance, power and energy details of each accelerator. We implement the micro-architectural pipeline of MindReading in FODLAM.

\section{Evaluation}
\begin{figure}[htbp!]
\vspace{-0.1in}
\centering
\includegraphics[width=3.4in]{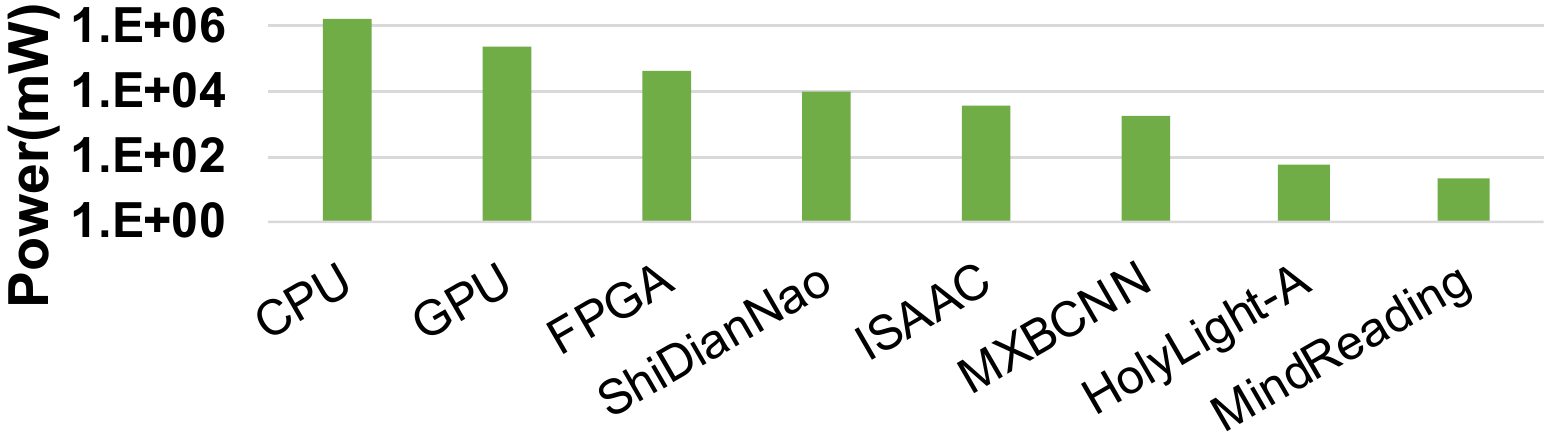}
\vspace{-0.1in}
\caption{Power consumption comparison.} 
\label{f:power_comparison}
\vspace{-0.1in}  
\end{figure}

\textbf{Power}. The comparison of power consumption of various accelerators is shown in Figure~\ref{f:power_comparison}. The ASIC-based ShiDianNao has less power consumption than CPU, GPU and FPGAs when processing 128 EEG-NET inferences per second since it is highly specialized for network inferences. The emerging ReRAM-based accelerator ISAAC reduces the power consumption by 59\% over ShiDianNao, because its ReRAM-based dot-product engines are more efficient. MXBCNN consumes less power than ISAAC when achieving 128-IPS, but has lower inference accuracy, due to its 4-bit binarized weights and activations. HolyLight-A significantly decreases the power consumption by 97\% over MXBCNN, since its photonic devices are highly power-efficent. However, it still requires 57.71 mW in which 79.1\% is consumed by a 16-bit adder and 256KB eDRAM. On the contrary, MindReading requires only a 4-bit adder and 64KB eDRAM. So it reduces the power consumption by 62.7\% over HolyLight-A.

\begin{figure}[htbp!]
\centering
\includegraphics[width=3.4in]{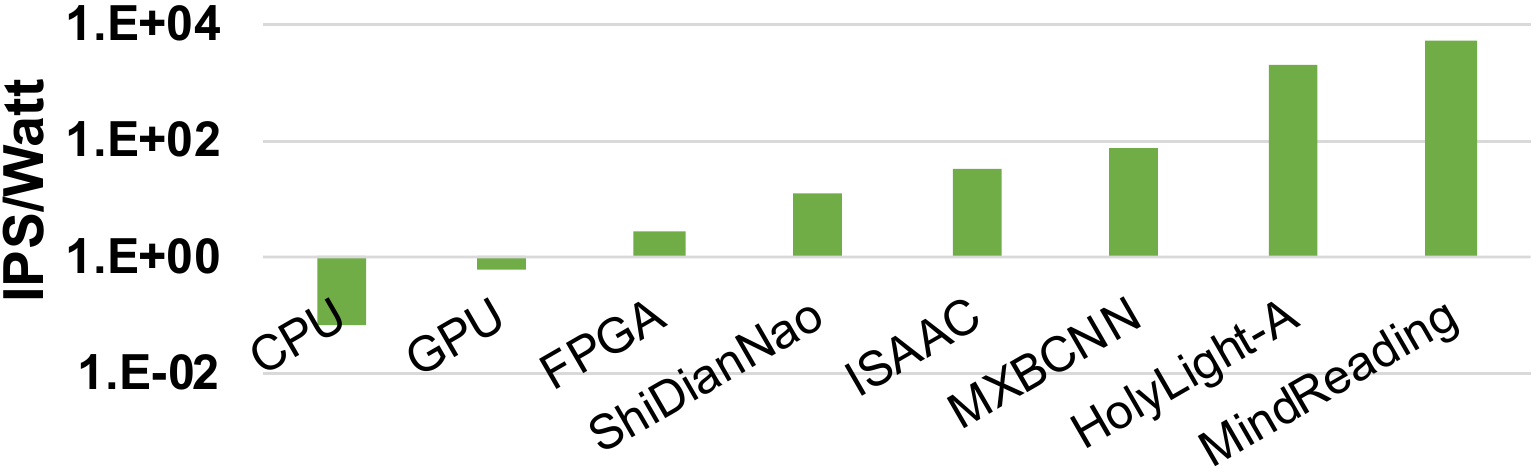}
\vspace{-0.1in}
\caption{Frames Per Second Per Watt comparison}. 
\label{f:fps_watt}
\vspace{-0.1in}  
\end{figure}

\textbf{Performance per Watt}.
The performance per Watt comparison of various accelerators is exhibited in Figure~\ref{f:fps_watt}. All non-photonic accelerators suffer from low performance per Watt. FPGA, CPU and GPU achieve only $<$ 5 IPS per Watt, while ShiDianNao, MXBCNN and ISAAC has $<$ 70 FPS per Watt. In contrast, the photonic accelerators, HolyLight-A and MindReading, boost the performance per Watt above 1000 IPS per Watt. Compared to HolyLight-A, MindReading improves the performance per Watt by 1.68$\times$, because it has less eDRAMs and lower bit-width photonic adder.

%Compared to MXBCNN, the emerging photonic accelerator, HolyLight-A, achieves 30.7 times improvement on FPS per Watt mainly using a 16-bit adder and 256KB eDRAM. MindReading beats HolyLight-A with a smaller adder and eDRAM to support LogP2QNN-base EEG-NET, achieving performance per Watt improvement over HolyLight-A.

%The EEG-NET performance per Watt comparison of various accelerators is exhibited in Figure~\ref{f:fps_watt}. FPGA and general-purpose computing units including CPU and GPU achieve $<$ 5 FPS per Watt. When the throughput of EEG-NET is required to 128 FPS, it is essential to use $>$ 25.6 Watt power. Sharing the similar trend to their power consumption, the performance per Watt of accelerators, e.g. ShiDianNao, Envision and ISAAC, achieve less than 50 FPS per Watt.  Compared to ISAAC, the emerging photonic accelerator, HolyLight-A, achieves 69.4 times improvement on FPS per Watt mainly using a 16-bit adder and 256KB eDRAM. MindReading beats HolyLight-A with a smaller adder and eDRAM to support LogP2QNN-base EEG-NET, achieving 3.8$\times$ performance per Watt improvement over HolyLight-A. 

\vspace{-0.1in}
\section{Conclusion}
In this paper, we present an ultra-low-power photonic accelerator, MindReading, to accelerate real-time human intention recognition. Compared to prior works, MindReading reduces the power consumption by 62.7\%, improves the throughput per Watt by 168\%, and meets the same real-time processing requirement.

\newpage
\bibliographystyle{IEEEtran}
\bibliography{brain}
\end{document}